\documentclass[useAMS,usenatbib,epsfig]{mn2e}
\usepackage{natbib}
\usepackage{amsmath,amsfonts}
\usepackage{epsfig}
\def\reff@jnl#1{{\rm#1\/}}

\def\aj{\reff@jnl{AJ}}                  % Astronomical Journal
\def\araa{\reff@jnl{ARA\&A}}            % Annual Review of Astron and Astrophys
\def\apj{\reff@jnl{ApJ}}                        % Astrophysical Journal
\def\apjl{\reff@jnl{ApJ}}               % Astrophysical Journal, Letters
\def\apjs{\reff@jnl{ApJS}}              % Astrophysical Journal, Supplement
\def\ao{\reff@jnl{Appl.Optics}}         % Applied Optics
\def\apss{\reff@jnl{Ap\&SS}}            % Astrophysics and Space Science
\def\aap{\reff@jnl{A\&A}}               % Astronomy and Astrophysics
\def\aapr{\reff@jnl{A\&A~Rev.}}         % Astronomy and Astrophysics Reviews
\def\aaps{\reff@jnl{A\&AS}}             % Astronomy and Astrophysics, Supplement
\def\azh{\reff@jnl{AZh}}                        % Astronomicheskii Zhurnal
\def\baas{\reff@jnl{BAAS}}              % Bulletin of the AAS
\def\jrasc{\reff@jnl{JRASC}}            % Journal of the RAS of Canada
\def\memras{\reff@jnl{MmRAS}}           % Memoirs of the RAS
\def\mnras{\reff@jnl{MNRAS}}            % Monthly Notices of the RAS
\def\pra{\reff@jnl{Phys.Rev.A}}         % Physical Review A: General Physics
\def\prb{\reff@jnl{Phys.Rev.B}}         % Physical Review B: Solid State
\def\prc{\reff@jnl{Phys.Rev.C}}         % Physical Review C
\def\prd{\reff@jnl{Phys.Rev.D}}         % Physical Review D
\def\prl{\reff@jnl{Phys.Rev.Lett}}      % Physical Review Letters
\def\pasp{\reff@jnl{PASP}}              % Publications of the ASP
\def\pasj{\reff@jnl{PASJ}}              % Publications of the ASJ
\def\qjras{\reff@jnl{QJRAS}}            % Quarterly Journal of the RAS
\def\skytel{\reff@jnl{S\&T}}            % Sky and Telescope
\def\solphys{\reff@jnl{Solar~Phys.}}    % Solar Physics
\def\sovast{\reff@jnl{Soviet~Ast.}}     % Soviet Astronomy
\def\ssr{\reff@jnl{Space~Sci.Rev.}}     % Space Science Reviews
\def\zap{\reff@jnl{ZAp}}                        % Zeitschrift fuer Astrophysik
\def\nat{\reff@jnl{Nature}}             % Nature 

\title[First results from the VSA -- IV. Cosmological 
parameter estimation]
{First results from the Very Small Array -- IV. Cosmological 
parameter estimation}
\label{firstpage}
\author[Rubi\~no-Martin et al.]  {Jose Alberto Rubi\~no-Martin$^1$,
  Rafael Rebolo$^{1,2}$, Pedro Carreira$^3,\dagger$, 
  Kieran Cleary$^3$, \newauthor
  Rod D. Davies$^3$, Richard J. Davis$^3$, Clive Dickinson$^3$, Keith
  Grainge$^4$, \newauthor Carlos M. Guti\'{e}rrez$^1$, Michael P. Hobson$^4$,
  Michael E.  Jones$^4$, R\"udiger Kneissl$^4$, \newauthor 
  Anthony Lasenby$^4$, Klaus Maisinger$^4$, Carolina \"{O}dman$^4$, Guy G.
  Pooley$^4$, \newauthor Pedro J. Sosa Molina$^1$, Ben
  Rusholme$^{4,\star}$, Richard D.E. Saunders$^4$, \newauthor 
  Richard Savage$^4$, Paul F. Scott$^4$, An\v{z}e Slosar$^4$, 
  Angela C. Taylor$^4$,
  \newauthor David Titterington$^4$, Elizabeth Waldram$^4$, 
  Robert A. Watson$^{3,\dagger}$ \newauthor and Althea Wilkinson$^3$ \\
  $^1$Instituto de Astrof\'{\i}sica de Canarias, 38200 La Laguna, Tenerife,
  Canary Islands\\
  $^2$Consejo Superior de Investigaciones Cient\'{\i}ficas, Spain \\
  $^3$Jodrell Bank Observatory, University of Manchester, UK\\
  $^4$Astrophysics Group, Cavendish Laboratory, University
  of Cambridge, UK\\
  $^\dagger$Present address: Instituto de Astrof\'{\i}sica de Canarias\\
  $^\star$Present address: Stanford University, Palo Alto, CA, USA}
\date{Accepted ---; received ---; in original form \today}
\pagerange{\pageref{firstpage}--\pageref{lastpage}}
\pubyear{2002}

\begin{document}
%\label{firstpage}
\maketitle

\begin{abstract}
We investigate the constraints on basic cosmological parameters set by
the first compact-configuration observations of the Very Small Array
(VSA), and other cosmological data sets, in
the standard inflationary $\Lambda$CDM model. 
Using the weak priors $40 < H_0 < 90$ 
km s$^{-1}$ Mpc$^{-1}$ and $0 < \tau < 0.5$, 
we find that the VSA and COBE-DMR data alone produce the
constraints $\Omega_{\rm tot} = 1.03 ^{+0.12} _{-0.12}$, 
$\Omega_{\rm b} h^2 = 0.029 ^{+0.009} _{-0.009}$, 
$\Omega_{\rm cdm} h^2 =  0.13 ^{+0.08} _{-0.05}$ and
$n_{\rm s} = 1.04 ^{+0.11} _{-0.08}$ at the 68 per cent confidence level. 
Adding in the type Ia supernovae constraints, we additionally find
$\Omega_{\rm m} = 0.32 ^{+0.09} _{-0.06}$ and 
$\Omega_\Lambda = 0.71 ^{+0.07} _{-0.07}$. 
These constraints are consistent 
with those found by the BOOMERanG, DASI and MAXIMA experiments. 
We also find that, 
by combining all these CMB experiments and assuming the 
HST key project limits for $H_0$ (for which the X-ray plus 
Sunyaev--Zel'dovich route gives a similar result), we obtain the tight 
constraints $\Omega_m=0.28^{+0.14}_{-0.07}$  
and  $\Omega_\Lambda= 0.72^{+0.07}_{-0.13}$, which are consistent
with, but independent of, those obtained using the supernovae data.
\end{abstract}

\begin{keywords}
cosmology: observations -- cosmic microwave background
\end{keywords}

\section{Introduction}

One of the central aims of cosmology is to determine the values of the
fundamental cosmological parameters that describe our Universe.
A unique opportunity to achieve this goal is provided by the 
observation of anisotropies in
the cosmic microwave background (CMB) radiation. 
By comparing such observations with the
predictions of our current theories of structure formation and the evolution of
the Universe, we may place constraints on 
the cosmological parameters that appear in these models.

The currently most favoured theoretical model for describing our Universe
is based on the idea of inflation \citep{guth-82}, which provides a natural 
mechanism for producing initial density fluctuations described by a power-law
spectrum with a slope close to unity. The simplest versions of 
inflation also predict the Universe to be spatially flat. 
The initial spectrum of adiabatic density fluctuations is modulated
through acoustic oscillations in the plasma phase prior to recombination
and the resulting inhomogeneities are 
then imprinted as anisotropies in the CMB. 
In the basic 
inflationary scenario, the CMB temperature anisotropies 
are predicted to follow a multivariate Gaussian distribution, and so
may be completely described in terms of their angular power spectrum.
Moreover, the acoustic oscillations in the plasma phase lead to a
characteristic series of harmonic peaks in the power spectrum, 
which are a robust
indicator of the existence of fluctuations on super-horizon scales.

Although the presence of acoustic peaks in the CMB power spectrum is
a generic prediction of inflationary models, detailed features of the 
power spectrum, such as the relative positions and heights of the
peaks, depend strongly on a wide range of cosmological
parameters, see e.g. \citet{hu-97}. Indeed, this sensitivity to the 
parameters is the reason why observations of the CMB provide
such a powerful means of constraining theoretical models.

Thus measurement of the CMB power spectrum is a major goal of observational
cosmology and numerous experiments have provided estimates of the spectrum on a
range of angular scales. It is only recently, however, that
observations by the BOOMERanG \citep{netterfield-01}, 
DASI \citep{halverson-02} and MAXIMA \citep{lee-01} experiments have provided
measurements of the CMB power spectrum with sufficient accuracy over a wide
range of scales to allow tight constraints to be placed on a wide range of
cosmological parameters (see, for example, \cite{du-pis}). 
This parameter estimation 
process is performed
by comparing the observed band-powers with a wide range of theoretical power
spectra corresponding to different sets of values of the cosmological
parameters, which can be accurately calculated using the {\sc Cmbfast}
\citep{zaldarriaga00} or {\sc Camb} \citep{al-camb} software packages. The
comparison of the observed and predicted power spectrum is usually carried out
in a Bayesian context by evaluating the likelihood function of the data as a
function of the cosmological parameters.

In this paper we perform this parameter estimation process using, as the main
CMB datasets, the flat band-power estimates of the CMB power spectrum 
measured by
the Very Small Array (VSA) in its compact configuration, which has been
described in the sequence of earlier papers \citet{VSApaperI}, 
\citet{VSApaperII} and \citet{VSApaperIII} (Papers I, II and III), 
and the COBE-DMR band-powers for low-$\ell$ normalisation.
We also combine the VSA data with other recent CMB experiments, and 
constraints from the HST Key Project on $H_0$ and observations of type
Ia supernovae, to tighten further the constraints on the cosmological
parameters. Two different
methods are used to perform the parameter estimation procedure. First, we
employ the traditional technique of evaluating the likelihood function on a
large grid in parameter space. Second, we consider a more flexible approach in
which the likelihood function is explored by Markov-Chain Monte Carlo (MCMC)
sampling.  The latter method has a great potential in terms of expanding the
dimension of the parameter set which can be investigated. Here we use it to
demonstrate the robustness of the results from the standard grid
approach and also to provide a 
novel visualisation of the range of uncertainty in our parameter
estimates.

\section{Models, parameter space and methods}\label{models}

In the analyses presented in this paper we restrict our attention
to the case in which the initial fluctuations are 
adiabatic with a simple power-law spectrum; such perturbations are
naturally produced in the standard single-field inflationary
model. Moreover, as is now common practice, we
consider models in which the contents of the Universe are divided into
three components: ordinary baryonic matter; cold dark matter (CDM), which
interacts with baryonic matter solely through its gravitational
effect; and an intrinsic vacuum energy. The 
present-day contributions of these
components, measured as a fraction of the critical density required to make the
Universe spatially flat, are denoted by $\Omega_{\rm b}$, $\Omega_{\rm
cdm}$ and $\Omega_\Lambda$ respectively.
It is possible that some of
the dark matter may, in fact, be in the form of 
relativistic neutrinos (hot dark
matter), but the presence of a hot component has a negligible effect on
the power spectrum, given the sensitivity and angular resolution of
current CMB experiments~\citep{dodelson96}. We therefore assume that all dark
matter is cold and set $\Omega_\nu=0$.

Following the current theoretical expectation~\citep{lyth97}, we
also assume that the contribution of tensor mode perturbations is very
small compared with the scalar fluctuations, and so we ignore their
effects. This assumption is consistent with current observations.
Since tensor modes contribute primarily to low multipoles, $\ell$,
the only existing measurement that would be particularly sensitive to their
presence is the level of the CMB power spectrum at low-$\ell$
observed by the COBE-DMR experiment~\citep{Smoot92}.
If the tensor component made up a large fraction of this
observed power, the value of the spectral index $n_s$ for scalar
perturbations would need to exceed unity by a
considerable margin in order to
provide the level of power at higher $\ell$ measured by 
numerous other CMB experiments. Such a large value of $n_s$ is,
however, ruled out by large-scale structure studies \citep{lss-ns}.
Nevertheless, it must be remembered that this argument only holds if
the initial perturbation spectrum is indeed a simple power-law.

Given the assumptions outlined above, there remain seven
degrees of freedom in the description of the standard inflationary CDM
model. The parameterisation of this seven-dimensional model space
can be performed in numerous ways, but we shall adopt the most common 
choice, which is defined by the following parameters: the physical
density of baryonic matter ($\Omega_{b}h^2\equiv \omega_{\rm b}$); 
the physical density of CDM ($\Omega_{\rm cdm}h^2\equiv \omega_{\rm
cdm}$); the vacuum energy density due to a cosmological constant
($\Omega_\Lambda$); the total density ($\Omega_{\rm tot}$); the
spectral index of the initial power-law spectrum of scalar perturbations
($n_{\rm s}$); the optical depth to the last-scattering
surface due to reionisation ($\tau$); and the overall normalisation of
the power spectrum as measured by $Q \equiv \sqrt{5C_2/(4\pi)}$ and is
quoted relative to $Q_{\rm COBE}$, as determined from the 4-year
COBE-DMR data by \cite{bunnwhite}.
This choice of parameters is similar to that made in the
analysis of the CMB band-power measurements obtained by the BOOMERanG,
MAXIMA and DASI experiments. 
We note that, in this parameterisation, the reduced Hubble
parameter $h$ ($\equiv H_0$ km s$^{-1}$ Mpc$^{-1}$/100) 
is auxiliary and is given by
$h=\sqrt{(\omega_{\rm b}+\omega_{\rm cdm})/(\Omega_{\rm
tot}-\Omega_\Lambda)}$.

In comparing the observed CMB flat band-powers measured by the VSA
with the above multidimensional model, we adopt a Bayesian
approach based on the evaluation of the likelihood function for the
data as a function of the cosmological parameters, which for brevity
we denote by the vector $\btheta = (\omega_{\rm b}, \omega_{\rm cdm}, 
\Omega_\Lambda, \Omega_{\rm tot}, n_{\rm s}, \tau, Q/Q_{\rm COBE})$.
To set proper
constraints on the values of these parameters, it is necessary to
explore the seven-dimensional model space over a region large enough
to encompass those models with significant likelihood.   
To that end, we employ two different techniques to explore the
likelihood function: a traditional approach in which the likelihood
function is evaluated on a grid of points in parameter space, and a
Markov-Chain Monte Carlo (MCMC) technique in which a set of samples
are drawn from the likelihood function. These techniques are described
below. A common requirement of both methods, however, is the accurate 
evaluation of the likelihood function at any given point in the 
parameter space, and we begin by describing how this computation is
performed.

\subsection{Evaluation of the likelihood function}\label{likeli}

As described in Paper III, the constraints imposed by the VSA compact
array observations on the CMB power spectrum have been obtained
using the {\sc Madcow} analysis package~\citep{klaus}. This provides
a one-dimensional likelihood distribution for each
flat band-power ${\cal C}_B$, conditioned on the values
of the other band-powers at the joint maximum of the likelihood
function. In addition the Hessian matrix at the joint maximum is also
calculated, which may be inverted to obtain the elements
$V_{BB'}$
of the covariance matrix
of the flat band-power estimates under the assumption that the
likelihood function is well approximated by a multivariate Gaussian near its
peak.

Since the VSA measures power only on small angular scales ($\ell >
150$), it does not constrain the amplitude and tilt of the power
spectrum at low-$\ell$, which leads to pronounced degeneracies in
the parameter space. Therefore, in Section~\ref{parms_vsa}, we also include
in our analysis the 28 COBE-DMR band-powers provided in
the {\sc Radpack} distribution~\citep{RADPACK}. Moreover, in
Section~\ref{combine},  
we further  
include CMB band-power measurements obtained by the BOOMERanG, MAXIMA
and DASI experiments.

In order to compare a set of observed flat band-powers ${\cal C}_{B,o}$ with
our theoretical model, it is necessary to compute the corresponding predicted
values ${\cal C}_{B,p}$ of these band-powers, 
given a particular set of parameter values $\btheta$. If $C_\ell$ is
the corresponding theoretical power spectrum for this set of parameter
values, the predicted value of the $B$th flat band-power is given by
\[
{\cal C}_{B,p} =\sum_\ell \frac{W_B(\ell)}{\ell} {\cal C}_\ell
%= \sum_\ell (2\ell+1) W_B(\ell)C_\ell/4\pi,
\]
where ${\cal C}_\ell = \ell(\ell+1)C_\ell/(2\pi)$ and
$W_{\rm B}(\ell)$ are the window functions for the
experiment under consideration. The VSA window functions are presented
in Paper III.

In the comparison of the observed and predicted band-powers it is necessary to
take proper account of the uncertainties in our estimates of
the ${\cal C}_{B,o}$. In fact, the uncertainties in band-power
estimates are, in general, non-Gaussian and this precludes us from
calculating a simple $\chi^2$-statistic. It is
possible, however,  to make a transformation from the flat
band-powers to a related set of `offset log-normal'
variables for which the uncertainties
are Gaussian to a very good approximation~\citep{bond00}. 
This requires the calculation of an additional set of quantities 
$x_B$ from the data, which represent the uncertainty due to
instrumental noise. These are straightforwardly calculated 
for the VSA band-powers, and will be available on our 
web-site\footnote{http://www.mrao.cam.ac.uk/telescopes/vsa}.
The corresponding quantities for the DMR and DASI band-powers are
included in the {\sc Radpack} package, and those for MAXIMA are
available on their website. For BOOMERanG, however,
the $x_B$ values are not yet publicly available, and so we assume
simply that the likelihood is a multivariate Gaussian. 

\begin{figure}
\epsfig{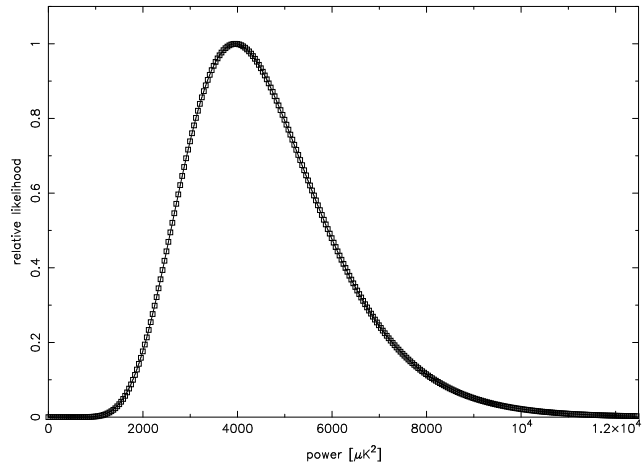}
\caption{An example of true likelihood (points) and its offset
log-normal approximation (line) for the flat band-power in the first VSA
spectral bin for the combined data from the VSA1, VSA2 and VSA3 fields.}
\label{fig:yadi}
\end{figure}

The observed band-powers are then transformed as
\[
\widetilde{\cal C}_{B,o}  =  \ln({\cal C}_{B,o}+ x_B),
\]
and similarly for the predicted band-powers.
It is straightforward to show that the elements of the covariance matrix
$\widetilde{V}_{BB'}$ for the new variables are related to the
covariance matrix $V_{BB'}$ of the original variables by
\[
\widetilde{V}^{-1}_{BB'} 
= ({\cal C}_{B,o}+x_B)(V)^{-1}_{BB'}({\cal C}_{B',o}+x_{B'}).
\]
The likelihood function is then taken to be a multivariate Gaussian
in the transformed variables, so that
\[
{\cal L}(\btheta) \propto \exp(-{\textstyle\frac{1}{2}}\chi^2),
\]
where the $\chi^2$ misfit statistic is given by
\[
\chi^2 = \sum_{B,B'}
(\widetilde{\cal C}_{B,o} - \widetilde{\cal C}_{B,p})
(\widetilde{V})^{-1}_{BB'}(\widetilde{\cal C}_{B',o} - 
\widetilde{\cal C}_{B',p}).
\]
We find that this assumed form for the likelihood function provides 
an excellent fit to the true likelihood distribution for the
VSA band-powers. This is illustrated 
in Fig.~\ref{fig:yadi}, in which we plot the
true likelihood and the corresponding
offset log-normal approximation for the flat band-power in the first
VSA spectral bin for the combined data from the VSA1, VSA2 and VSA3 fields.

We note that the use of this transformation not only allows us to find the
best-fit point in the parameter space $\btheta$, but also allows us
to use the value of $\chi^2$ at this point as a measure of goodness of
fit.

\subsection{Exploration of the parameter space: 
grid-based approach}\label{grid}

Once one has the facility to calculate accurately the 
likelihood function ${\cal L}(\btheta)$ for the
cosmological parameters $\btheta$ at any given point in the parameter
space, one must devise a strategy to explore the likelihood
distribution throughout this space. 

From the above discussion, it is clear that the evaluation of the
likelihood function at each point in parameter space requires one to
calculate the theoretical power spectrum corresponding to that point.
The calculation of this $C_\ell$ spectrum is performed using version
4.0 of the {\sc Cmbfast} code \citep{zaldarriaga00}, 
which uses a $k$-space splitting
technique that accelerates the calculation of the spectrum by using
a flat $\Omega_{\rm tot}=1$ model to compute the high-$\ell$
multipoles for models with $\Omega_{\rm tot} \neq 1$. Nevertheless,
the calculation of a typical spectrum for a $\Omega_{\rm tot} \neq 1$
model still requires around 30 sec of CPU time on one of 
the processors of the
{\sc Cosmos} SGI Origin 2000 computer. Even with the 16 processors
available to us, this computational cost severely limits the
total number of points in parameter space at which the likelihood
function can reasonably be evaluated. We note that, in fact, two
complete grids of models (and two independent pipelines for the
cosmological parameters estimation) were set up in Tenerife and Cambridge.

The traditional approach is thus to calculate the $C_\ell$ spectra
for a grid of models that is as fine as CPU-time limitations allows, 
while being sufficiently large to encompass the entire 
probability distribution. For our model space, we calculated $C_\ell$ spectra
on a six-dimensional grid corresponding to the values 
for each parameter given in Table~\ref{gridtable}. 
%
%\begin{center}
\begin{table*}
\caption{The values of the parameters in the six dimensional grid of
models.\label{gridtable}}
\begin{tabular}{llllllllllllll}
\hline\hline
$\Omega_{\rm b} h^2$:    & 0.010 & 0.015 & 0.020 & 0.025 & 0.030 & 0.035 
 & 0.040 & 0.045 & 0.050 \\
$\Omega_{\rm cdm} h^2$: & 0.02  & 0.06  & 0.1   & 0.2   & 0.3   &  0.4
 & 0.5 & 0.6 & 0.7 & 0.8 & 0.9 & 1.0 & 1.1 \\ 
$\Omega_\Lambda$: &  0.0 & 0.1 & 0.2 & 0.3 & 0.4 & 0.5 & 0.6 & 0.7 & 0.8\\
$\Omega_{\rm tot}$: & 0.7 & 0.75 & 0.80 & 0.85 & 0.90 & 1.00 &
1.05 & 1.10 & 1.15 & 1.20 & 1.30 \\
$n_s$: & 0.7 & 0.8 & 0.9 & 1.0 & 1.1 & 1.2 & 1.3 & 1.4 \\
$\tau$:  & 0.0 & 0.025 & 0.05 & 0.075 & 0.1 & 0.2 & 0.3 & 0.5 \\
%$Q/Q_{\rm COBE}$: & 0.45 & 0.56 & 0.67 & 0.78 & 0.89 & 1.00 & 1.11 &
%1.22 & 1.33 & 1.44 
\hline\hline
\end{tabular}
\end{table*}
%\end{center}
%
%All this grid was computed using a Beowulf cluster of 10 computers.
The six-dimensional grid contains $419\,968$ spectra. Since the
different values of the scalar spectral index $n_{\rm s}$ can all be 
obtained on a single call to {\sc Cmbfast}, this required $52\,496$ runs in
total and took $\sim 2$ days of CPU time on {\sc Cosmos}.
The overall
normalisation parameter $Q/Q_{\rm COBE}$ need not be precomputed
on a grid since it produces a simple linear scaling of the $C_\ell$
spectra. In this parameter, the likelihood function was calculated
at 10 points in the range 0.45 to 1.44, with a step size of 0.11.

The likelihood function ${\cal L}(\btheta)$ for a given dataset 
can be evaluated at each of these grid points as 
described in section~\ref{likeli}, and the location of the maximum determined.
For each cosmological parameter, the
one-dimensional marginalised distribution is then obtained by
numerically integrating over the other parameters. These marginal
distributions are then interpolated with a cubic spline, and determine
the constraints placed on the cosmological parameters.

Wherever possible, we also include the calibration uncertainties of the
CMB experiments under consideration as extra parameters in our
analysis. The prior on such parameters is taken as a Gaussian centred
on unity, with a standard deviation of the appropriate width.  An
analytic marginalisation is performed over this parameter, using the
method proposed by \citet{bridle-marg}. This 
analytical procedure assumes, in addition to a Gaussian prior on the
calibration parameter, that the likelihood function is
Gaussian. Unfortunately, in neither the original band-powers, 
nor the offset log-normal variables, are {\em both} these
functions precisely Gaussian, and so some (small) approximation is introduced.
In this paper, the analytical marginalisation is
performed before transforming to offset log-normal variables.

\subsection{Exploration of the parameter space: MCMC approach}\label{mcmc}

In a parameter space of large dimensionality, a natural alternative
to the grid-based approach is instead to sample from the likelihood
distribution. An efficient procedure for obtaining such samples is
to construct a Markov chain whose equilibrium distribution corresponds to
the likelihood function in parameter space (see e.g. \citet{knox-age}). Thus
after propagating the Markov chain for a given
`burn-in' period, one obtains samples from the likelihood
distribution, provided the chain has converged.

The MCMC sampling procedure may be implemented most 
straightforwardly by using a simplified version of the 
Metropolis--Hastings algorithm.  At each
step $n$ in the chain, the next state $\btheta_{n+1}$ is chosen by
first sampling a candidate point $\btheta'$ from some proposal
distribution $\pi(\btheta)$. The candidate point is then accepted with
probability 
\[
\alpha = \mbox{min}\left[1,\frac{\pi(\btheta_n){\cal P}(\btheta_{n+1})}
{\pi(\btheta_{n+1}){\cal P}(\btheta_n)}\right].
\]
where ${\cal P}(\btheta)$ is the posterior distribution 
for the parameters $\btheta$ and is simply
the product of the likelihood ${\cal L}(\btheta)$ and
and some prior $p(\btheta)$; we take the latter to be uniform unless
otherwise stated.
If the candidate point is accepted, the next state in the chain
becomes $\btheta_{n+1}=\btheta'$, but if the candidate point is
rejected, the chain does not move, and $\btheta_{n+1}=\btheta_n$.
In theory, the convergence of the chain to the limiting 
distribution is independent of the choice of the proposal distribution
but this choice is crucial in determining both the 
rate of convergence to the limiting distribution and
the efficiency of the subsequent sampling. 
An effective approach
is to set the proposal distribution $\pi(\btheta)$ to a multivariate
Gaussian, centred on the current point in parameter space.

As mentioned above, the states of the chain $\btheta_n$ can be
regarded as samples from the limiting (i.e. likelihood) distribution
only after some initial burn-in period for the chain to reach
equilibrium. The topic of convergence is still a matter of statistical
research and no definitive formula exists for calculating the
required length of the burn-in period. Nevertheless, several
convergence diagnostics have been proposed \citep{burn-in}, which may be used
as a guide. 

After burn-in, the sample density is directly proportional to the 
likelihood distribution, and so the samples may be used
straightforwardly to obtain estimates of the parameter values and
confidence limits. In particular, one may easily obtain
one-dimensional marginalised distributions for each parameter
separately, obviating the need for computer-intensive numerical
integrations. Moreover, the computational requirements of 
MCMC procedures are almost independent of dimensionality of parameter
space and thus allow a large number of parameters to be
constrained simultaneously by the data. 

The strength of MCMC methods lies in the fact that useful parameter
estimations can be achieved with considerably fewer
likelihood evaluations as compared to the traditional grid-based
methods. Nevertheless, the density of samples must be high enough so
that the estimation of the underlying posterior probability
distribution is not plagued by Poisson noise and is independent of
the kernel density estimation method. Usually one 
requires on the order of several tens of thousands of (accepted) samples.
The efficiency with which these samples may be obtained depends 
strongly on the shape of the posterior likelihood function and in
practice the basic Metropolis--Hastings algorithm can be augmented by
the use of various speed-ups. Most notably, the sampling efficiency
is improved by considering several simultaneous
correlated Markov chains and by specific random point generators that
attempt to follow the shape of the likelihood distribution posterior.
The main MCMC implementation used here is that provided by the {\sc Bayesys}
software (Skilling, private communication), which employs several
enhancements of the basic Metropolis--Hastings algorithm and has the
ability to sample using multiple chains. 

Ideally, one would like to calculate a theoretical power spectrum
using one of the popular packages, such as {\sc Camb} or
{\sc Cmbfast} for each sample. However, this is
computationally extremely expensive. As a practical alternative, 
one can use a pre-computed grid of theoretical spectra, 
such as that discussed in section~\ref{grid}, and 
calculate the required spectrum by suitable interpolation between
neighbouring grid points. The density of grid points in the parameter
space must be small enough that the dominant error in the estimated
cosmological parameters comes from the errors in the measured CMB
power spectrum and not from the interpolation between the grid points.
We tested the accuracy of the grid discussed in section~\ref{grid} 
by calculating the exact CMB power
spectrum using {\sc Cmbfast} for 1000 randomly chosen sets of parameters
and  comparing them 
with grid interpolation. We find that the rms error on the predicted
band-powers resulting from the interpolation
is around 4 per cent, which is very small as compared with the
uncertainties in the band-powers from the current CMB experiments.

The {\sc Bayesys} sampler provides a powerful general purpose
MCMC engine, which allows one to explore complicated likelihood functions
of large dimensionality. We find, however, that the present accuracy
of the CMB experiments results in  likelihood surfaces that are relatively 
smooth and highly convex, which can be adequately explored using a
simple MCMC sampler based solely on the straightforward
Metropolis--Hastings algorithm. This allows the possibility
of tailoring an 
MCMC algorithm to the specific problem of cosmological parameter
estimation from CMB band-power measurements, and taking advantage of
our prior knowledge concerning which parameter combinations can be
calculated quickly using {\sc Cmbfast} or {\sc Camb} and which
directions in parameter space to avoid. Such an approach has been
implemented in the software package {\sc Cosmomc} \citep{lewisbrid}, 
and allows one to perform an MCMC exploration of the
parameter space in which the use of a grid is bypassed altogether, and
the exact theoretical $C_\ell$ spectrum is calculated at each sample
point. We have made use of this approach as an additional cross-check of our
results.

In addition to using the MCMC approach to provide 
useful checks of the parameter constraints
obtained from the standard grid-based method, we
have exploited the insensitivity of the MCMC method to the 
dimensionality of the
problem by including calibration uncertainty in our numerical
analysis, rather than performing an approximate analytical
marginalisation over it, as performed in the grid-based approach.
A new parameter $a$ is introduced which is the ratio of the \emph{real} 
telescope calibration to the experimental best estimate. 
A Gaussian prior is assumed for $a$, with its centre at unity and with a
standard deviation corresponding to the calibration uncertainty of
the experiment under consideration.
Whenever the MCMC algorithm requires a sample for a given value of
$a$, the data are rescaled accordingly.

\section{Cosmological parameter constraints from the VSA}\label{parms_vsa}

We first consider the constraints placed on the values of the 
cosmological parameters 
$\btheta = (\omega_{\rm b}, \omega_{\rm cdm}, 
\Omega_\Lambda, \Omega_{\rm tot}, n_{\rm s}, \tau, Q/Q_{\rm COBE})$
using only the VSA band-powers, 
and the low-$\ell$ normalisation provided by the 28 COBE-DMR band-powers
provided in the RADPACK package. The precise VSA data used
were the $10$ band-power estimates and associated covariance matrices
for each of the three separate 3-field mosaics VSA1, VSA2 and VSA3. 

\subsection{Grid-based approach}\label{vsa_grid}
\begin{figure*}
\epsfig{file=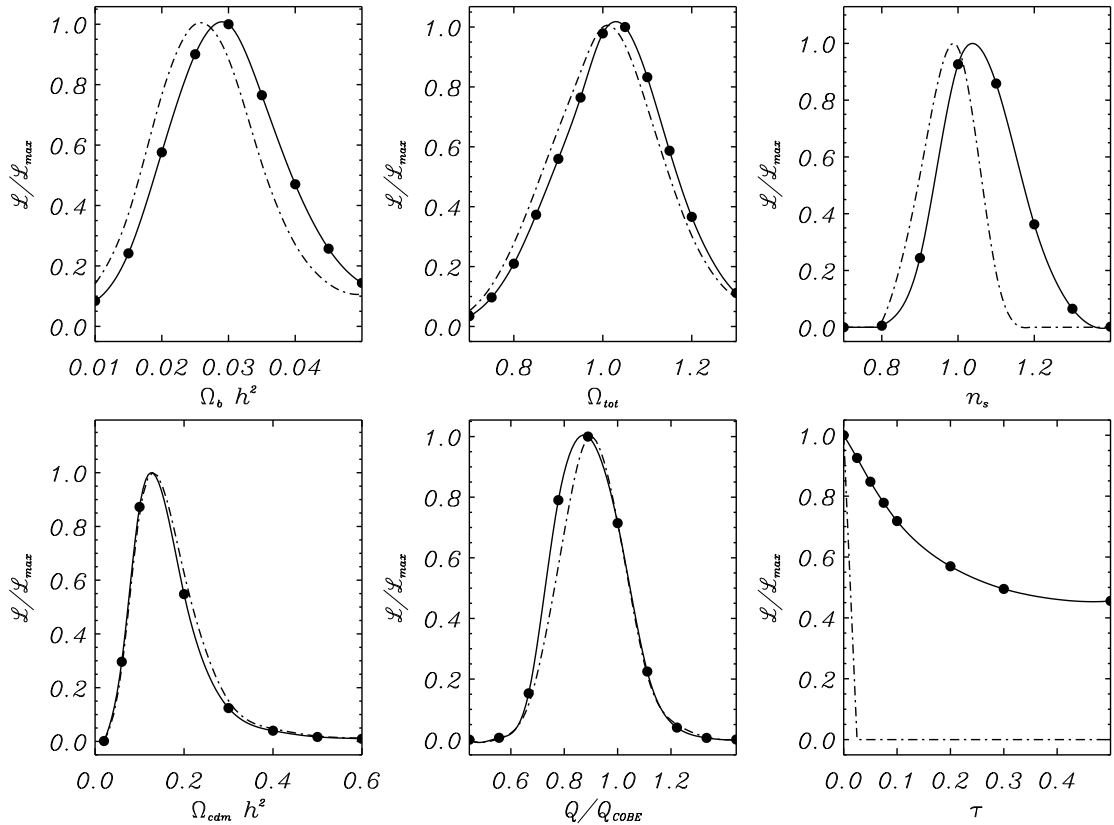,width=11cm,angle=90}
\caption{Marginalised distributions for the cosmological parameters 
as determined from the VSA experiment, together with a normalisation
constraint at low-$\ell$ from COBE-DMR. The solid lines 
include the effect of a weak top-hat prior on $h$ $(0.4 < h < 0.9)$
and an implicit top-hat prior on $\tau$ $(0<\tau < 0.5)$ resulting
from the grid boundaries. The dot-dashed lines correspond 
to the additional prior $\tau=0$.
The lines are constructed using a cubic spline interpolation through
the grid points, which are shown as solid circles.}
\label{fig:vsa_params}
\end{figure*}
\begin{table*}
\caption {Cosmological parameters estimated from VSA and COBE-DMR data,
using several priors. It should be noted that these estimates include
the effect of the implicit priors on the parameters resulting from the
finite grid of models (see Table~\ref{gridtable}). 
All the confidence regions correspond to 
the 68 per cent level.}
\label{tab:vsaiac}

\begin{tabular}{|c|c|c|c|c|c|c|} 
\hline 
\hline
Prior &  $\Omega_{b} h^2$& $n_{s}$ & $\Omega_{tot}$ & 
$\Omega_{cdm} h^2$ & $\Omega_m$ & $\Omega_{\Lambda}$\\
\hline 
\{$0.4 < h < 0.9$\} &  $0.029 ^{+0.009} _{-0.009}$ & 
	$1.04 ^{+0.11} _{-0.08}$  & $ 1.03 ^{+0.12} _{-0.12}$ &
	$0.13 ^{+0.08} _{-0.05}$  & - & - \\
\{$0.4 < h < 0.9$\} + \{$\tau = 0$\} &  $0.026 ^{+0.008} _{-0.008}$ &
	$0.99 ^{+0.06} _{-0.07}$   & $1.01 ^{+0.12} _{-0.13}$ &
	$0.13 ^{+0.08} _{-0.05}$ & - & -\\
\{$10\,\mbox{Gyr} < \mbox{age} < 20\,\mbox{Gyr}$\} 
&  $0.028 ^{+0.009} _{-0.008}$& 
	$1.02 ^{+0.11} _{-0.08}$  & $1.05 ^{+0.14} _{-0.12}$ &
	$0.12 ^{+0.05} _{-0.04}$  & - & - \\
\{$10\,\mbox{Gyr} < \mbox{age} < 20\,\mbox{Gyr}$\} 
+ \{$\tau = 0$\} & $0.025 ^{+0.008} _{-0.008}$ &
	$0.97 ^{+0.06} _{-0.07}$   & $1.03 ^{+0.15} _{-0.12}$ &
	$0.12 ^{+0.06} _{-0.04}$ & - & -\\
\{$h= 0.72 \pm 0.08 $\} + \{$10\,\mbox{Gyr} < \mbox{age} < 
20\,\mbox{Gyr}$\} & $0.028 ^{+0.007} _{-0.008}$ & 
	$1.00 ^{+0.06} _{-0.05}$ & $0.96 ^{+0.06} _{-0.12}$ & 
	$0.19 ^{+0.08} _{-0.07}$ & $0.48 ^{+0.08} _{-0.21}$ & 
	$0.47 ^{+0.33} _{-0.16}$ \\
\hline
\{SNIa\} + \{$0.4 < h < 0.9$\}  &  $0.029 ^{+0.009} _{-0.009}$ & 
	$1.02 ^{+0.12} _{-0.08}$ & $1.02 ^{+0.08} _{-0.06}$  & 
	$0.09 ^{+0.05} _{-0.04} $ & $0.32 ^{+0.09} _{-0.06}$ & 
	$0.71 ^{+0.07} _{-0.07}$ \\
\hline
\hline 
\end{tabular}
\end{table*}

Using the approach outlined in section~\ref{grid}, we calculate the
corresponding likelihood function ${\cal L}(\btheta)$ over the
six-dimensional grid summarised in Table~\ref{tab:vsaiac} and the 
normalisation parameter $Q/Q_{\rm COBE}$. In addition, we include 
the calibration uncertainty of the VSA band-powers as an extra
parameter in our analysis. The prior on this parameter is taken as a
Gaussian centred on unity, with a standard deviation corresponding
to the known calibration error for the VSA of 7 per cent in $(\Delta T)^2$. 

After analytically 
marginalising over calibration uncertainty, the best-fit model
is characterised by the parameter values 
$(\omega_{\rm b}, \omega_{\rm cdm}, 
\Omega_\Lambda, \Omega_{\rm tot}, n_{\rm s}, \tau, Q/Q_{\rm COBE})
=(0.020, 0.06, 0.7, 1.0, 0.9, 0.0, 0.87)$, but no particular
significance should be attached to this model. 
It is the marginalised constraints on the individual parameters 
that are most important. Nevertheless, it is of interest to determine
the goodness-of-fit for this model. At the peak, the value of $\chi^2$
was found to be $47.3$ for the $3\times 10$ VSA plus 28 COBE-DMR
band-powers. Assuming that a full 7 degrees of freedom are lost to the
fit (which is unlikely given the form of the theoretical power
spectra), we thus have 51 remaining degrees of freedom. 
The value $\chi_{51}^2=47.3$ lies at the 38 per cent point
on the $\chi_{51}^2$ cumulative distribution function, which is entirely
acceptable and shows that model to be a good fit to the data.

For a multidimensional likelihood function calculated
on a grid, one has already implicitly assumed top-hat priors on each of the
parameters, corresponding to the edges of the grid. It should
also be bourne in mind that, in principle, if the grid does not
encompass all of the
likelihood in {\em any} parameter, then that top-hat prior becomes
relevant for {\em all} of the parameters. In addition to the implicit
prior arising from the grid, we may also impose explicit 
priors on the 
cosmological parameters values, according to our existing knowledge 
(or prejudices) concerning their values. In particular, we consider
combinations of the following five priors: (i) a weak top-hat prior on 
$h$ ($0.4 < h < 0.9$); (ii) a strong prior Gaussian on $h$ 
($h=0.72\pm 0.08$, \citet{hst-final}); (iii) a weak top-hat prior on age 
($10\,\mbox{Gyr}\,< t < 20\,\mbox{Gyr}$); (iv) a strong prior on
optical depth ($\tau=0$); (v) a prior in the 
$(\Omega_{\rm m},\Omega_\Lambda)$-plane from high-redshift Type IA
supernovae (where $\Omega_{\rm m}=\Omega_{\rm b}+\Omega_{\rm cdm}$)
\citep{perlmutter99}. After adopting (combinations of) these priors, we
then obtain the one-dimensional marginalised distribution for each
cosmological parameter by direct numerical integration; the
successive integrals over the parameter directions are evaluated in turn
by first performing a cubic spline interpolation onto a fine 
regularly-space grid of points.

An illustration of our results is shown in Fig.~\ref{fig:vsa_params}
for the two cases in which we assume prior (i) above on $h$, 
with and without the additional prior (iv) on $\tau$.
The solid lines represent results treating $\tau$ as a free parameter,
whereas the dot-dashed lines correspond to setting $\tau=0$.
It is clear from the figure that the effect of this latter prior is minimal,
leading only to minor changes in the constraints
on $\Omega_{b} h^2$ and $n_{s}$, which is to be
expected from the well-known degeneracy between 
these two parameters \citep{ns-obh}. 

The 68 per cent confidence limits derived from these marginal
distributions on the
cosmological parameters $\omega_b$, $\omega_{\rm cdm}$, 
$\Omega_{\rm tot}$ and $n_{\rm s}$ are given in the first two rows 
of Table~\ref{tab:vsaiac}. The upper and lower limit in each case is 
defined such that the corresponding interval contains 68 per cent of the total
probability, and the likelihood function evaluated at
each limit has the same value; the quoted best-fit value is the mode
of the corresponding marginalised distribution.
Also listed in the table are the 68 per
cent confidence 
limits on these parameters resulting from (combinations of) the 
different priors listed above. We note that the constraints
on these four parameters are relatively insensitive to the inclusion
of increasingly stringent priors. In particular, we see that
the constraints on $\omega_{\rm b}$ are all consistent with the
constraint $\omega_{\rm b}=0.020\pm 0.002$
$(95\%)$~\citep{burles-01} resulting from the observed
primordial hydrogen-deuterium ratio and the theory of
nucleosynthesis.  Depending on the prior assumed, the 
preferred value of $\omega_{\rm cdm}$ is typically around 0.12, 
with an uncertainty of 
0.05, and lends very strong support to existing evidence for the 
existence of non-baryonic dark matter. We also note that, 
for all the priors considered, the constraints on the total density 
$\Omega_{\rm tot}$ are consistent with the Universe being
spatially flat. Finally, and again independently of the particular prior
assumed, the constraints on the scalar spectral index are consistent
with the scale-invariant $n_{\rm s}=1$ (Harrison--Zel'dovich) initial
power spectrum, which is preferred by standard inflationary models.

In the first four rows of Table~\ref{tab:vsaiac},
the priors are unable to break the well-known degeneracy of CMB data
in the $(\Omega_{\rm m},\Omega_\Lambda)$-plane, and so no
constraints are given. Nevertheless, with the assumption of our
strong prior on $h$ and a weak prior on the age of the Universe,
we see that one begins to break this degeneracy. Indeed, one finds that
the preferred values of $\Omega_{\rm m}$ and $\Omega_\Lambda$ correspond to
a roughly equal partitioning of the critical density between matter (baryonic
and dark) and vacuum energy, although the lower 
limit for $\Omega_m$ and the upper limit for 
$\Omega_\Lambda$ extend some way from the best-fit values. 
Finally, the assumption of the 
high-redshift SNIa constraint \citep{perlmutter99} 
and our weak prior on $h$, succeed in cleanly
breaking the CMB degeneracy and we find the partitioning of the
critical density between matter and vacuum energy is well constrained 
in the approximate proportions one-third to two-thirds.

\subsection{MCMC approach}\label{vsa_mcmc}

Using precisely the same data as analysed in the previous section,
we also explored the seven-dimensional parameter space $\btheta$
using the MCMC techniques outline in section~\ref{mcmc}. First, we used
the {\sc Bayesys} algorithm and, for each sample, calculated the
theoretical $C_\ell$ spectrum by interpolating from the pre-computed 
grid. Since the shape of the 
likelihood function is very simple, it was enough to 
run just eight simultaneous Markov chains, with each walk
requiring only a very short burn-in period. After burn-in, we
collected 3000 samples from each chain, thus obtaining 24000 samples
in total from the likelihood distribution. As one would hope, we find
that all parameter constraints calculated from these samples are
fully consistent with those obtained above for each set of imposed
priors. We therefore do not
reproduce them here, although they do provide a useful check on our
earlier results.

\begin{figure}
\epsfig{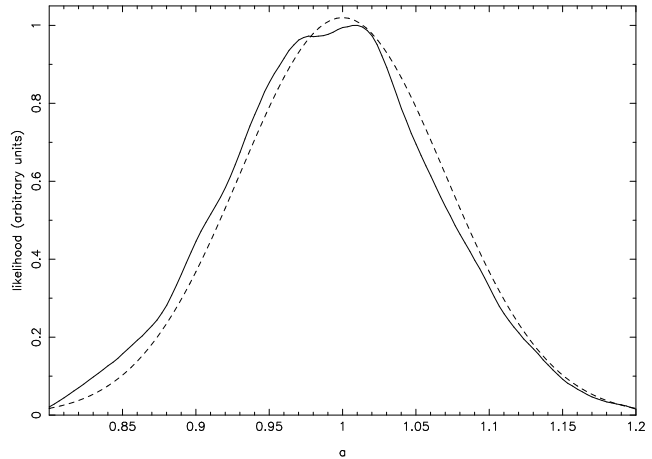}
\caption{The marginalised distribution for the calibration parameter 
$a$ after analysing the data (solid line). Also plotted is the
original prior placed on $a$ (dashed line).}
\label{fig:a}
\end{figure}

As mentioned in section~\ref{mcmc}, to illustrate the flexibility of the
MCMC approach, we included the calibration uncertainty of the VSA
band-power measurements as an additional parameter in our numerical
analysis. The prior on this parameter $a$ was taken to be
a Gaussian centred at unity, with standard deviation $0.07$. 
In Fig.~\ref{fig:a}, we plot this prior distribution, together with the
marginalised distribution on the parameter $a$ {\em after} analysing
the data. The mean of this  distribution lies at $a=1.00$ and has a
standard deviation of 0.071. Thus, we see that, within the class of
models considered, the measured CMB band-powers are consistent with
our estimated calibration uncertainty.

\begin{figure}
\epsfig{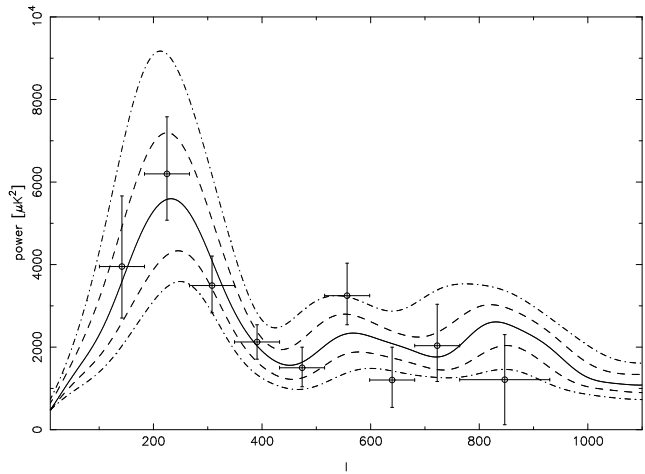}
\caption{An illustration of the MCMC sample density on the
$(l,C_l)$-plane. The solid
line represents the maximum of the sample density at a given $\ell$-value, 
while dashed and dot-dashed lines
correspond to 68 per cent and 95 per cent confidence limits. 
The vertical lines are VSA
data points with 68 per cent confidence limits. 
Note that samples were obtained using
full covariance information, which cannot be displayed 
conveniently on a single plot.}
\label{fig:b}
\end{figure}

As a second illustration of the usefulness of the MCMC approach, 
we plot in Fig.~\ref{fig:b} a novel representation of the
sets of cosmological models consistent with the VSA plus COBE-DMR 
data, which is
produced as follows. 
Each of the MCMC samples corresponds to a theoretical
$C_\ell$ spectrum. Thus from the samples it is straightforward 
to construct a one-dimensional distribution of the power at each
multipole $\ell$. In Fig.~\ref{fig:b}, the position of the 
maximum of the distribution at each $\ell$ is shown by the solid line,
while the dashed and dot-dashed lines indicate the
68 and 95 per cent confidence limits of the distribution
respectively, determined in the same manner as in section~\ref{vsa_grid}.
This plot assumes our earlier weak prior on the age of
the Universe. We note that the first peak is very well defined, and
that there is also good evidence for the second peak. 

As mentioned in section~\ref{mcmc}, we also check our cosmological
parameter constraints by using the straightforward {\sc Cosmomc}
algorithm, which is optimised for the problem at hand and bypasses the
need for a grid. Once again our results are fully consistent with
those found above, and provide another useful check on our analysis.

\section{Combining VSA with other CMB experiments}\label{combine}

So far we have only combined the VSA data with the COBE-DMR experiment
in order to place limits on cosmological parameters from the CMB.
It is clear,
however, that tighter constraints may be obtained if we additionally include
information from other CMB experiments, in particular
BOOMERanG~\citep{bernardis-peaks}, DASI~\citep{cosmo-params-from-dasi}, 
and MAXIMA~\citep{stompor-01}. 

\begin{figure}
\begin{center}
\epsfig{file=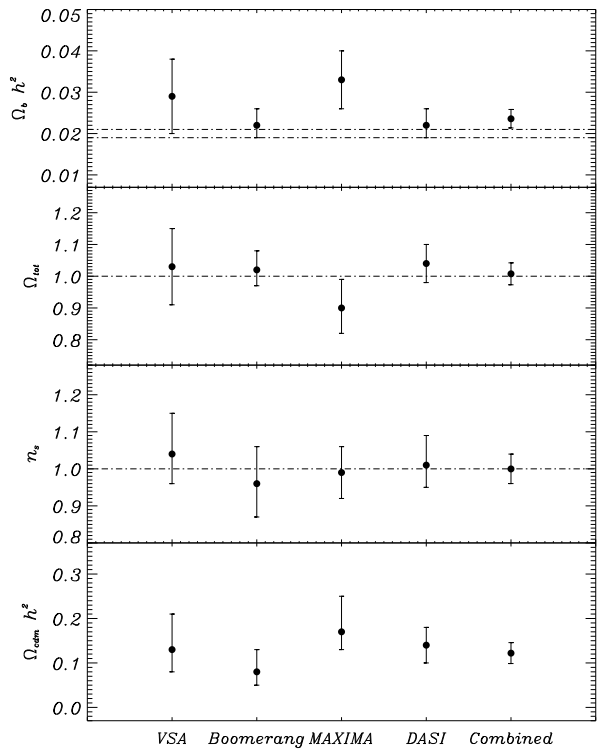,width=\linewidth}
\caption{Comparison of the maximum likelihood estimates and
68 per cent confidence limits on the 
cosmological parameters 
$\Omega_b h^2$, $\Omega_{tot}$ and $n_{s}$ 
from the VSA and recent other CMB experiments (see text).}
\label{fig:compare1}
\end{center}
\end{figure}

In Fig.~\ref{fig:compare1},  we begin by simply plotting the maximum-likelihood
estimates for $\Omega_{\rm b} h^2$, $\Omega_{\rm tot}$, $n_{\rm s}$ and
$\Omega_{\rm cdm} h^2$
obtained by the VSA and these other recent CMB experiments, together
with the reported 68 per cent confidence intervals 
(except for MAXIMA, for which we plot as error bars $1/2$ of the
reported 95 per cent confidence limits). We note that the confidence
limits for each experiment include the effects of calibration and
beam uncertainty, where appropriate, and each experiment also assumes
the COBE-DMR band-powers and similar weak priors
to those adopted in section~\ref{vsa_grid}.

In general, the
constraints on each individual parameter
agree within error bars. Moreover, 
since each of these experiments uses different observing
techniques and has observed different regions of the sky, 
we may consider each measurement as an independent estimate of the
corresponding parameter. Assuming further that the individual
likelihoods are Gaussian, they may be immediately combined to produce
a joint constraint on each parameter, which is also shown in the
figure.

In the top panel, corresponding to the parameter $\Omega_b h^2$, 
we also plot the 68 per cent confidence limits arising
from the nucleosynthesis constraint
$\Omega_b h^2 = 0.020 \pm 0.001$ \citep{burles-01}.
The combined measurement from all CMB experiments is consistent with the BBN
constraint and fully supports the case for a low value of the primordial
deuterium abundance. It also favours a primordial helium mass fraction of 
$Y_p \approx 0.246$. In the second panel, we see that all the experiments 
agree with the prediction $\Omega_{tot}=1$ of standard inflationary
models. Remarkably, 
the combined CMB measurement has an error bar of only $\sim$ 3 per cent.
The combined value for $n_{s}$  also agrees at the 1-$\sigma$  level
with the scale-invariant Harrison-Zel'dovich initial power spectrum,
i.e. with $n_{\rm s}=1$. Finally, we see that the combined 
value $\Omega_{\rm cdm}h^2$ is tightly constrained to be around 5 times
larger than the value for $\Omega_{\rm b}h^2$, which is a strong
indication for the existence of non-baryonic matter.
%with the results from MAXIMA ( $\Omega_{cdm} h^2$=0.2$^{+0.1}_{-0.05}$) 
%and DASI ( $\Omega_{cdm} h^2$=0.14$\pm$0.04).

\begin{figure}
\begin{center}
\epsfig{file=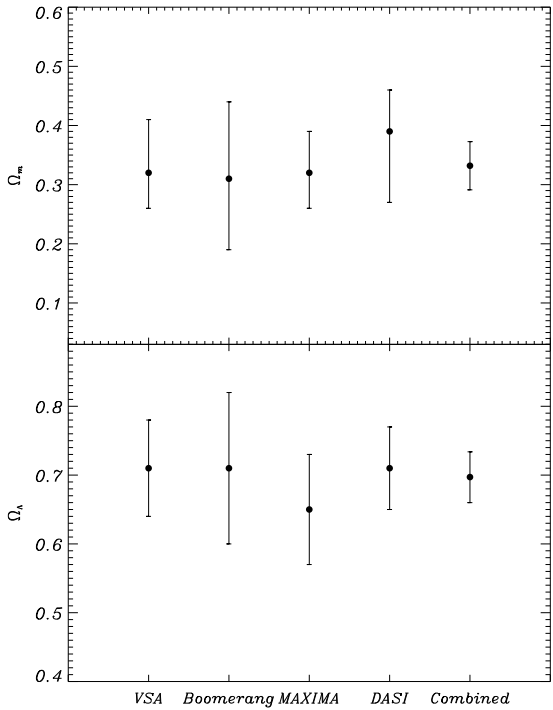,width=8cm}
\caption{Comparison of the maximum likelihood estimates for $\Omega_m$ and
$\Omega_\Lambda$ from the VSA and recent CMB experiments. A prior 
from SNIa measurements is adopted (Perlmutter et al. 1999), as well as a weak prior on $H_0$.}
\label{fig:compare2}
\end{center}
\end{figure}

In Fig.~\ref{fig:compare2} we compare the VSA+SNIa constraints 
on $\Omega_{\rm m}$ and $\Omega_\Lambda$, with those published  
by the other recent CMB experiments. It is important to note, however,
that the published DASI value does not make use of any SNIa data, so in order
to enable a proper comparison to be made, we have repeated the
complete grid-based parameter estimation procedure for DASI
performed by \citet{cosmo-params-from-dasi}, using their
published covariance matrices
and window functions, but
including the SNIa prior from \citet{perlmutter99} and a weak prior on
$h$. Incidentally, we found that the parameter constraints we
derived for DASI before including the SNIa prior were in complete
agreement with those obtained by \citet{cosmo-params-from-dasi}.
In Fig.~\ref{fig:compare2}, we plot the 68 per cent confidence limits 
from each experiment, and we see that they are all in good agreement. 
The combined constraint is $\Omega_{\rm m}=0.33 ^{+0.04}_{-0.04}$  
and  $\Omega_\Lambda= 0.70^{+0.04}_{-0.04}$.

As noted previously by \cite{cosmo-params-from-dasi},
by combining all the available CMB
datasets together with a strong prior on $h$, it is 
possible to break cleanly the CMB
degeneracy in the $(\Omega_{\rm m},\Omega_\Lambda)$-plane {\em
without} using the SNIa prior. 
As our CMB datasets, we use
the VSA  and COBE-DMR band-powers, 
the RADPACK compilation for DASI, plus the published results 
from BOOMERanG and MAXIMA. 
By performing a simple grid-based parameter estimation
procedure, as outlined in section~\ref{grid}, we find that one can break
the  $(\Omega_{\rm m},\Omega_\Lambda)$-plane CMB degeneracy by
assuming only our weak prior on the age of the Universe 
(10 Gyr $<$ age $<$ 20 Gyr) and our strong Gaussian prior on $h$ 
$(h=0.72 \pm 0.08)$ from the HST key project. Our resulting
constraints are $\Omega_m=0.28^{+0.14}_{-0.07}$  
and  $\Omega_\Lambda= 0.72^{+0.07}_{-0.13}$, which are similar to
those derived above, but are independent of the SNIa data.
Hence, this result is not subject to the usual uncertainties
associated with using high-redshift type IA supernovae as standard
candles. We stress that a very similar result would be obtained by
instead combining the CMB data with the prior on $h$ from 
Sunyaev--Zel'dovich and X-ray observations of clusters~\citep{jones-02}. 

\section{Discussion}\label{discussion}

%In this paper we have presented results of cosmological parameter
%estimation using the newly available data from the VSA compact array 
%experiment, and have also combined these new data with other existing 
%recent CMB experiments and constraints from the Type IA supernovae data.

In Figs \ref{fig:compare1} and \ref{fig:compare2} we have presented
results of cosmological parameter extraction from various CMB
experiments. These experiments use a variety of observational
techniques and operate at a range of frequencies and have therefore
widely different systematic effects. Nevertheless, the figures show 
remarkable agreement between different experiments. 
This may indicate the importance of the assumed weak priors
(which are often common) and possibly the fitting of too many parameters
given the constraining power of individual experiments. Indeed, the
reduced $\chi^2$ is below $1$ for most experiments (see individual papers). 
Nevertheless, when one combines experiments (VSA, DASI,
BOOMERanG, MAXIMA and DMR data) most cosmological parameters become
constrained at the level ranging between 5--20 per cent, if one
neglects the possibility of tensor modes.
This accuracy
rivals the discriminatory power of the upcoming CMB satellite experiments,
such as MAP \citep{map}, with the added bonus that the residual
systematic effects of the various experiments are diluted.

We also verify that a constraint on
the vacuum energy component of the Universe may be obtained 
{\em independently} of the Type Ia supernovae data,
by combining the results from CMB experiments alone, together with the
a prior on $h$ from the HST key project \citep{hst-final}. 
Moreover, the resulting
values for $\Omega_\Lambda$ and $\Omega_{\rm m}$ are consistent with
those obtained when supernovae data are included.
It is often assumed that CMB experiments cannot
constrain the vacuum energy, as a result 
of the well known CMB degeneracy in the
$(\Omega_m,\Omega_\Lambda)$-plane, 
and the SNIa data are usually employed to break this
degeneracy. The supernovae data are, however, still somewhat plagued
by uncertain systematic effects (such as extinction along the line of sight and
the evolutionary effects due to metallicity), although the issues are
gradually being resolved \citep{sn-resolv}. Therefore, obtaining
independent consistent results on the value of the cosmological constant
increases our confidence in both the supernovae and CMB results.

\section{Conclusions}\label{conclusions}

In our analysis of the newly available data from the VSA compact
array, and other recent cosmological results, we have found the following.
\begin{itemize}
\item{Traditional grid-based methods and Markov-Chain Monte Carlo
(MCMC) sampling techniques have been applied to the cosmological
parameter estimation problem and found to yield consistent results.}
\item{The VSA observations, when combined with the COBE-DMR data and a
weak top-hat prior on $h$ ($0.4 < h < 0.9$) give
$\Omega_{\rm tot} = 1.03 ^{+0.12} _{-0.12}$, 
$\Omega_{\rm b} h^2 = 0.029 ^{+0.009} _{-0.009}$, 
$\Omega_{\rm cdm} h^2 =  0.13 ^{+0.08} _{-0.05}$ and
$n_{\rm s} = 1.04 ^{+0.11} _{-0.08}$ at the 68 per cent confidence
level.}
\item{Adding in observations of type Ia
supernovae, the CMB degeneracy in the 
$(\Omega_{\rm m},\Omega_\Lambda)$-plane may be broken to yield
the constraints $\Omega_{\rm m} = 0.32 ^{+0.09} _{-0.06}$ and 
$\Omega_\Lambda = 0.71 ^{+0.07} _{-0.07}$.} 
\item{The BOOMERanG, DASI and MAXIMA experiments, which have different
approaches and systematics, yield consistent constraints on
cosmological parameters, which is gratifying. Combining the results
of these recent CMB experiments with the VSA data, and
assuming weaks priors on $h$ and the age of the Universe, gives 
$\Omega_{\rm tot} = 1.01 ^{+0.03} _{-0.03}$, 
$\Omega_{\rm b} h^2 = 0.024 ^{+0.002} _{-0.002}$, 
$\Omega_{\rm cdm} h^2 =  0.12 ^{+0.02} _{-0.02}$ and
$n_{\rm s} = 1.00 ^{+0.04} _{-0.04}$.}
\item{Adding in the type Ia supernovae constraint to the combined
CMB result, the $(\Omega_{\rm m},\Omega_\Lambda)$-plane degeneracy
is cleanly broken to give $\Omega_m=0.33 ^{+0.04}_{-0.04}$  
and  $\Omega_\Lambda= 0.70^{+0.04}_{-0.04}$.}
\item{One can equally well break the 
$(\Omega_{\rm m},\Omega_\Lambda)$-plane degeneracy without the SNIa
data, by assuming the strong prior on $h$ from either the HST Key
Project or Sunyaev--Zel'dovich and X-ray observations of clusters.
This yields the constraints 
$\Omega_{\rm m}=0.28^{+0.14}_{-0.07}$  
and  $\Omega_\Lambda= 0.72^{+0.07}_{-0.13}$.}
\end{itemize}

\section*{ACKNOWLEDGEMENTS} 

We thank Sarah Bridle for providing her grid-based likelihood
programs, Antony Lewis for making available his 
{\sc Cosmomc} code and John Skilling for 
allowing us to use his {\sc Bayesys} MCMC sampler. 
This research was conducted in cooperation with SGI utilising the
HEFCE-supported COSMOS supercomputer. We also acknowledge the IAC Computer
Centre for helping with the setup of a 10-processor Beowulf system
dedicated to this research.
We thank the staff of the Mullard Radio Astronomy Observatory,
Jodrell Bank Observatory and the
Teide Observatory for invaluable assistance in the commissioning and operation
of the VSA. The VSA is supported by PPARC and the IAC. Partial
financial support was provided by the Spanish Ministry of Science and
Technology project AYA2001-1657. A. Taylor, R. Savage, B. Rusholme and 
C. Dickinson acknowledge support by PPARC studentships. K. Cleary
and J.A. Rubi\~no-Martin acknowledge Marie Curie Fellowships 
of the European Community programme
EARASTARGAL, 'The Evolution of Stars and Galaxies', under contract
HPMT-CT-2000-00132. K. Maisinger acknowledges a Marie Curie Fellowship
of the European Community. A. Slosar acknowledges the support of
St. Johns College, Cambridge. We thank Professor Jasper Wall for
assistance and advice throughout the project.

\label{lastpage}
\bibliography{paper4_final}
\bibliographystyle{mn2e}
\bsp

\end{document}